# Controls and Interfaces

*Q. King*
CERN, Geneva, Switzerland

**Abstract**
Reliable powering of accelerator magnets requires reliable power converters and controls, able to meet the powering specifications in the long term. In this paper, some of the issues that will challenge a power converter controls engineer are discussed.

**Keywords**
Power converters; regulation; control; reliability.

## 1    Introduction

Power converter control has moved more and more into the digital domain. As a result, the choices of control hardware have evolved extremely rapidly, and the potential performance of that hardware has expanded exponentially. Exploiting the potential of the hardware has fallen on software and programmable logic developers, with the result that the effort needed to develop converter control software can now surpass the effort needed to develop the hardware.

The diversity of potential hardware solutions means that it is not possible to proscribe a 'right way' to approach the control of converters. Instead, this short paper introduces some of the topics that should be considered to meet both performance and reliability targets. It will focus in particular on the control of continuously regulated power converters, rather than fast pulsed converters.

## 2    Defining the requirements

It is obvious that finding a good controls solution starts with defining the problems to be solved. This can be surprisingly difficult. At its most basic, the powering of an accelerator will involve a number of circuits, mostly involving magnets, but some may be providing high voltage for RF klystrons or other RF devices, or electrostatic elements for low-energy beams. Obviously the power engineers need to know the required rating of each converter, but for the controls a lot of additional information should be captured for each circuit, including the following.

– What will be the regulated signal: voltage, current or field?

– How is the reference of this value defined?

     o    DC (e.g. for storage ring or Linac circuits).

     o    Function of time run on demand (e.g. for non-cycling accelerators such as the Large Hadron Collider (LHC)).

     o    Function of time run by a cyclic timing system (e.g. for cycling accelerators such as the CERN Proton Synchrotron (PS)).

     o    Calculation in real-time based on an outer regulation loop (e.g. orbit or tune).

– How accurately must the value be controlled? It is important to distinguish between absolute accuracy, reproducibility, and stability. These are treated in more detail in Ref. [1].

–   What are the rate of change and acceleration limits?

From this you can define the requirements for the timing system, the analogue measurement system, and the regulation. This is not simple and will take time. Ultimately, this will lead to the key requirements of the controller hardware, including:

–   the regulation rate;

–   the measurement rate;

–   the processing power;

–   the noise and resolution of the analogue acquisition system;

–   the interface for sending the reference (voltage and/or modulation) to the power converter.

## 2.1   Scalability

A very important factor in the design of the controls is the scale of the system. If you have ten circuits then an elaborate automatic configuration management system is not justified. Once there are 100 circuits, managing them individually will start to become time-consuming; and once a system has more than 1000 circuits, then automatic management tools are mandatory.

The calibration of the analogue measurement components will be important, particularly for the main circuits of an accelerator, which are usually the most demanding of accuracy. If an analogue component such as an analogue to digital converter (ADC) or direct current–current transformer (DCCT) is replaced, a method is needed to ensure that the control of the circuit can continue to meet the specification after the intervention. This can be manual for a small system, but for large installations an automatic calibration system is highly desirable.

## 3      Converter controls reliability

It is vital to remember that getting all of the circuits to work according to the specification when the facility is commissioned is only part of the challenge. It is equally important to keep them working with the required reliability for the lifetime of the facility. For this challenge, the mean time between failures (MTBF), mean time to repair (MTTR), and the scale of the system are all important. The bigger the system, the more attention should be given to reducing the MTBF and MTTR.

## 3.1   Hardware reliability

It is obvious that the MTBF of the global system will depend on the MTBF of the controller hardware. This is too big a subject to treat in detail in this short paper, but experience shows that once early failures have been resolved, control electronics can achieve remarkably high levels of reliability. One million hours MTBF per controller has been achieved at CERN; however, to get to this level requires attention to every aspect of the design and production (and some luck). For more information about practical reliability, consult the Bibliography. Here are some points to keep in mind.

–   Use only the best quality connectors.

–   Avoid the need for wiring by using circuit boards to link circuits.

–   Design for test and, if you expect to manufacture more than a hundred or so units, build test hardware at the same time. Base your test hardware on standard off-the-shelf components, such as PXI cards. If your device is hard to test then it is probably too complicated and should be redesigned to be more modular.

–   Avoid fans if possible, while keeping the operating temperature of components below 50°C. If fans have to be used, consider using a temperature-based controller so they only run as fast as necessary. Choose fans with the highest MTBF you can get and try to mount them so that the axis of rotation is vertical. Monitor the temperature in the controller and have a warning threshold. Consider preventative maintenance by replacing all of the fans once they have reached around 70–80% of their rated operating life.

– Pay close attention to electromagnetic compatibility (EMC). Follow best practice for grounding and carry out burst tests to see if the required immunity has been achieved. Use relays and opto-couplers or optical fibres for long-distance signals, and design for the worst-case over-voltages on converter-related signals in the event of an earth fault on the magnet circuit.

– Use standard protocols, cables, and connectors.

– Avoid radiation areas if at all possible. If you have to design electronics to work in radiation, then allow a lot more time and money for their development and get expert advice. See Ref. [2] for more information about this.

– Avoid potentiometers and other adjustable components.

– Avoid electrolytic capacitors, and overate passive components for power and voltage.

– Don't miniaturize unnecessarily. Use the largest passive devices that fit and the biggest pad sizes on active devices. Only use ball grid array (BGA) packages when unavoidable. Include pads to mount test connectors for a logic analyser for use during the development phase.

– Exploit programmable logic and make it easy to update. For large installations, allow the logic to be reprogrammed over the network.

– Don't use programmable digital circuits just for the sake of it. If a function can be done simply with an analogue circuit, then this will need less effort to maintain. All software (including programmable logic) has a major overhead for maintenance in the long term.

– If the controller incorporates multiple circuit boards, enclose the assembly in a metal cassette.

– Plan for the obsolescence of the hardware and the development systems. How will you compile the logic and software in 25 years? Investigate virtual machines before it is too late.

Apart from good design, the operating conditions can impact reliability.

– Do not trust commercial power supplies. Either design your own or use a pair of supplies with monitored redundancy, or both.

– If possible, keep the electronics powered and warm all the time, reducing the thermal cycles to which the hardware is subjected.

– Keep powered spare controllers in the vicinity of the operational systems.

– Keep the operational environment clean of dust, with temperature and humidity close to nominal, and protect the electronics from water damage in the event of leaks from water-cooled cables or components in the power converter.

– Use halogen-free cables to reduce the impact of fire damage.

– Incorporate a means to remotely identify, and hence track, individual parts.

– Predict the expected failure rate of components, and put in place the means to track failures and repairs. This should be used to determine real-world failure rates and to identify new failure modes, allowing preventive actions to be taken in the case of unexpectedly high failure rates.

## 3.2  Converter spares policy

The power engineers will need to address the same issues of MTBF and MTTR when designing or specifying the power converters and when defining the spares policies. Different approaches may be taken according to the types of converters used. Below are two examples that have an impact on the specification for the controls.

### 3.2.1  *Converters made from modular components*

In this approach, large converters are made from multiple sub-converter racks containing multiple standard modules. The modules are light enough that one or two people can swap a faulty unit with a spare within a few minutes. In this case, a circuit can only be powered by one converter, but most failures will be in the power modules which can be quickly swapped. The controller will only ever be responsible for the one circuit, so addressing and configuration are static.

In this case, the challenge for the control system is to accurately identify which module needs to be replaced in the event of a fault. For this, a very reliable system to capture the first fault must be deployed. In the case of the LHC power converters, the power engineers adopted an $n + 1$ redundancy policy for this class of converters, so they all have one more sub-converter than those required to deliver the nominal current. The low-level converter electronics can compensate for the loss of a sub-converter in real time by increasing the current supplied by the others.

However, sometimes this may not work and the trip of a sub-converter may result in a cascade of trips in the other sub-converters. The LHC ATLAS experiment's toroidal field magnet converter has eight sub-converters, each containing six modules. If one module is unreliable and trips off randomly every day or so, and if this results in tripping the whole converter, then this will quickly become a problem unless the controls can accurately identify the faulty module.

This means that the first-fault logging must not only identify the first fault within a sub-converter, but also the first sub-converter to trip. This may require time-stamping of the faults with a resolution of the order of 10 µs.

### 3.2.2  *Monolithic converters*

For large monolithic converters, the power engineers may install one spare converter in an area, to cover a number of operational converters. In the event of a failure, the circuit cables are patched to the spare unit, which allows the faulty converter to be repaired later.

The spare converter might have its own controller, in which case it must take over the address of the controller of the failed converter. This can present some interesting configuration challenges if the spare converter is not exactly of the same type as the failed unit. Alternatively, the controller from the failed unit might be patched or moved to control the spare.

Whichever approach is adopted, the probability of mistakes by the team making the intervention should be minimized by reducing the complexity of their task.

## 3.3  Addressing the controller

The controller addressing scheme will depend upon the network architecture. However, it is reasonable to assume that the controller will be made from components that can be exchanged in the event of failure. So this raises the question of how a controller knows its address. This might be a MAC address on Ethernet, or a fieldbus address if a fieldbus is used.

The new controller's address could be configured manually by the team who are performing the intervention, perhaps by using jumpers or switches. But this is error prone and can easily result in two controllers appearing on the network with the same address. This almost inevitably causes disruption to the system and can be hard to diagnose, so there is a strong motivation to avoid this kind of error.

A preferred approach is to encode the address in some passive device that gets plugged onto the controller. This might be the network connector or it could be in a separate dongle. In either case, a simple circuit board can encode a digital value with copper tracks so that, once soldered, a failure of the dongle is highly improbable. Alternatively, an I²C electrically erasable programmable read-only memory (EEPROM) (or equivalent) could be used, but being an active device it will have a lower MTBF.

### 3.4    Analogue calibration

The conversion of an analogue signal into a scaled value inside the controller requires two distinct pieces of information:

–    the nominal scale factor;

–    the calibration error.

There is no avoiding the need for the nominal scale factor but different approaches are possible for the calibration error. In one approach, the analogue hardware can be designed with calibration components such as potentiometers. In this way, the calibration errors can be reduced to within the specification and the software can consider the measurement to be perfect.

While this simplifies the software, it puts a significant burden on the team responsible for keeping the analogue measurement devices calibrated and, in the case of potentiometers, it increases the chances of component failure and error.

A preferred approach is to avoid all adjustable components and to accept that all of the analogue measurement devices will have calibration errors. Provided that these errors can be measured, then they can be compensated in software. If such an approach is adopted, it also opens the possibility to compensate non-linearity and temperature effects.

If a known and trusted reference signal can be selected automatically then the software can measure the calibration errors itself. If not, then calibration of an analogue measurement device such as a DCCT might be done on a test stand. The measured calibration errors can then be stored for later use by the controller.

Where to store this calibration data is a key question. For a few circuits this could be entirely manual with the values written in a notebook and entered into the controller by hand. Obviously this is impractical for a big system with hundreds or thousands of circuits. An alternative is to store the calibration data in a non-volatile memory inside the device. This is a simple concept and can be effective for small- to medium-sized systems.

For large systems, using a central database is recommended. In this case, a way to identify the measurement device is needed. Ideally this should be machine-readable so that the controller can automatically identify the connected devices and request their calibration data from the database. Although this is a significant investment in terms of software development, it is an important step towards the goal of full automatic configuration.

### 3.5    Automatic configuration

For large systems, accurate automatic configuration of the controllers is hugely important for the reliability of the whole system. It is a worthy objective to avoid all manually configurable components such as switches and jumpers in the design of the hardware.

All configuration parameters can be stored centrally in a database. These parameters fall into three categories.

– Parameters associated with an individual component. For example, the calibration errors for a particular DCCT or ADC.

– Parameters associated with a type of a component. For example, the scale factor for a type of DCCT head.

– Parameters associated with an individual circuit. For example, the magnet inductance.

For the first category, individual components must be identified by the component type and a unique serial number. This is typically converted into a barcode that is stuck onto the component. For automatic configuration to be possible, the controller needs to be able to know the barcode of each attached component that has configuration parameters.

For the second category, the component type field from the barcode can be used to look up configuration parameters in the database associated with that type of component.

For the third category, the name of the circuit can be used to look up the circuit parameters. Obviously the control system needs to know the association between controller address and circuit name.

## 3.6   Software reliability

Converter control is a real-time problem, so many modern programming languages are unsuitable. The natural choice remains the low-level languages C and C++. Increasing processing power is making the real-time challenge easier to face, but bear in mind that embedded programming is time-consuming so, where possible, move the programming to a level where standard tools such as Linux (with real-time extensions) can be used. Please consult the Bibliography for recommended books on embedded systems programming.

It is obvious that software reliability is improved by adopting:

– a source code versioning tool such as git;

– a continuous integration tool such as Bamboo or Jenkins;

– an issue tracking tool such as JIRA or Trac;

– code reviews.

Unit testing may also help, but it has limits when programming real-time multi-threaded software.

With real-time programming, time is at the heart of everything, so invest in a way to record the time taken to execute every important part of the real-time code. This can be a simple min/max pair, or a histogram if the profile is significant.

# 4      Regulation

The core business of converter controls is (usually) the delivery of the correct current in the circuit as a function of time. In special cases, the field in a magnet might be measured and regulated, or the voltage provided to power an RF generator or an electrostatic element in a beam line.

Numerous control strategies are possible, but two that have been widely adopted for accelerator power converter control are:

– digital implementation of a classical 'analogue' regulator such as a proportional integral derivative (PID) controller using state equations;

– digital implementation of a digital regulator based on the RST polynomial algorithm.

Each has benefits. The classical approach is simple for a non-expert to tune as the number of parameters is small, while the RST approach can provide excellent performance with a constant tracking delay (the time between the reference being set and the measurement arriving at the reference value).

## 4.1 Classical regulation

Many laboratories have developed digital controllers that have implemented PID regulators using classical state equations. In particular, the Swiss Light Source at PSI has developed both electronics and corrector power supplies that have been widely used at many light sources [3]. The digital signal processor (DSP) in the electronics implements the current regulator and pulse width modulation (PWM) without an intermediate voltage regulation loop. This has allowed a high bandwidth of more than 1 kHz for the corrector circuits, which has been valuable for the orbit feedback needed by light sources to stabilize the electron beam in the insertion devices.

## 4.2 The RST algorithm

The RST algorithm is becoming increasingly popular because of its performance and flexibility. The RST algorithm is defined in Eq. (1),

$$\sum_0^n\{Act_i \cdot S_i\} = \sum_0^n\{Ref_i \cdot T_i\} - \sum_0^n\{Meas_i \cdot R_i\}\,, \tag{1}$$

where $i = 0$ corresponds to the current sample, $i = 1$ is the previous sample and so on. This notation was proposed by Landau [4]; however in many textbooks the $R$ and $S$ polynomials are exchanged. The application of the RST algorithm to power converter control is described in Ref. [5].

From Eq. (1), it is easy to see that if you know the new reference and measurement, you can calculate the new actuation, $Act_0$, if you keep the history of the previous $n$ samples of the reference, measurement, and actuation,

$$Act_0 = \frac{\sum_0^n\{Ref_i \cdot T_i\} - \sum_0^n\{Meas_i \cdot R_i\} - \sum_1^n\{Act_i \cdot S_i\}}{S_0}\,. \tag{2}$$

Typically the actuation will be the voltage reference for an inner voltage regulation loop, but if the DC bus is very stable and the PWMs are very linear, then actuation could be the modulation reference directly. Either way, if the actuation is limited, it is easy to back-calculate the reference that, when combined with the new measurement, will result in this limited actuation when used in Eq. (2). This is given in Eq. (3),

$$Ref_0 = \frac{\sum_0^n\{Act_i \cdot S_i\} + \sum_0^n\{Meas_i \cdot R_i\} - \sum_1^n\{Ref_i \cdot T_i\}}{T_0}\,. \tag{3}$$

In this way the reference history can be kept coherent with the measurement and actuation histories. This is equivalent to the anti-windup feature of a traditional regulation algorithm.

The benefit of the RST equation is that any linear regulator up to order $n$ can be implemented by choosing the appropriate RST polynomial coefficients. Simple PI, PID, or PII controllers can be implemented as well as more complex higher order systems, without changing the software. The challenge with the RST approach is the calculation of the coefficients. This cannot be done by hand and either requires an expert using MATLAB (or equivalent), or a library that encodes the knowledge of an expert for a particular type of load.

## 4.3 Circuit load model

The majority of the circuits in a typical accelerator are inductive and resistive. Figure 1 shows a generic first-order model that can cover most cases.

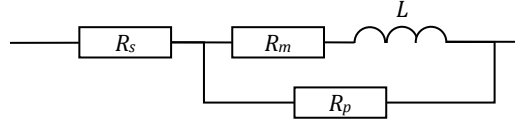

**Fig. 1:** Generic first-order inductive load model

There are three resistances in the model.

$R_s$ – series resistance representing the resistance of the cables.

$R_m$ – magnet resistance. This will be zero for superconducting magnets or circuits that do not contain a magnet.

$R_p$ – parallel resistance. This is rarely used but in some circuits in which many magnets are connected in series, the parallel resistance is needed to damp out resonances.

The transfer function for this first-order model is presented in Eq. (4),

$$G(s) = \frac{1}{R_s + \frac{1}{\frac{1}{R_p} + \frac{1}{R_m + sL}}} \,. \tag{4}$$

The frequency response of the gain (current/voltage) has the classic first-order form presented in Fig. 2.

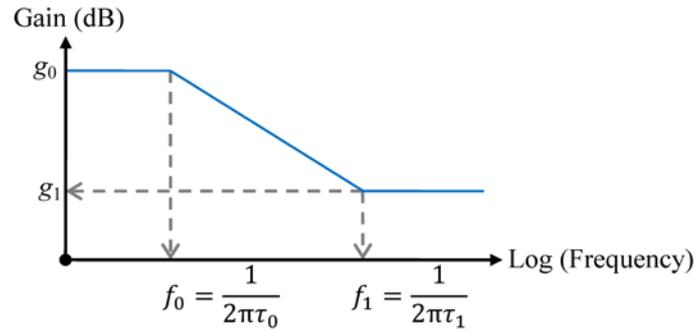

**Fig. 2:** Bode diagram for first-order load model

This response is defined by DC and high-frequency gains, $g_0$ and $g_1$, which can be calculated using Eqs. (5) and (6),

$$g_0 = \frac{1}{R_s + \frac{R_p R_m}{R_p + R_m}} \,, \tag{5}$$

$$g_1 = \frac{1}{R_p + R_s} \,. \tag{6}$$

The frequencies of the first-order pole and zero, $f_0$ and $f_1$, are defined by the periods $\tau_0$ and $\tau_1$, which can be calculated using Eqs. (7) and (8),

$$\tau_0 = \frac{L}{R_m + \frac{R_p R_s}{R_p + R_s}} \,, \tag{7}$$

$$\tau_1 = \frac{L}{R_p + R_m} \,. \tag{8}$$

For very large super-conducting magnets, the stray capacitance between coils can be significant (as discussed in the next section) and a different load model may be more appropriate.

### 4.3.1    Parallel resistance

If a circuit requires the parallel resistance to damp resonances, this may have an important consequence upon the control of the current in the magnet. The parallel resistor will allow some of the circuit current to bypass the inductor during transients.

Figure 3 illustrates the effect on the circuit current of this prompt response to steps in the voltage. It means is that for some time after each step in voltage, the magnet current will not equal the circuit current, which can be significant because it is the circuit current that the controller measures and regulates. It is important for the accelerator physicists to be aware of this fact, since they are interested in the magnet current.

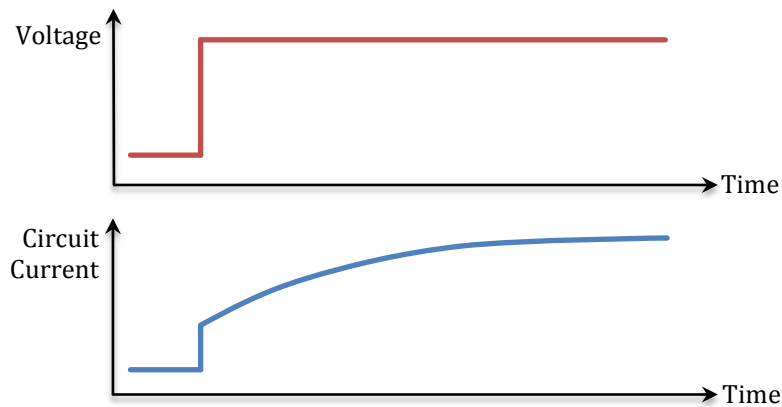

**Fig. 3:** The presence of a parallel resistance ($R_p$) in the load results in a prompt response in the current to steps in the applied voltage.

In the LHC, the main dipole and quadrupole circuits are the only ones that require parallel damping resistors. The characteristics of these circuits are given in Table 1. The key ratio is $g_1/g_0$, which are ~$10^{-7}$ and ~$10^{-6}$, respectively. These circuits need to be regulated with ppm level accuracy, so it is fortunate that this ratio is very small. A lower value of $R_p$ would have jeopardized the quality of the regulation of the magnet current.

**Table 1:** LHC main circuit characteristics

| Circuit | $L$ | $R_s$ | $R_p$ | $R_m$ | $g_0$ | $g_1$ | $\tau_0$ | $\tau_1$ | $f_0$ | $f_1$ |
|---------|-----|-------|-------|-------|-------|-------|----------|----------|-------|-------|
| Dipoles | 15.7 | $1\times10^{-3}$ | $1.54\times10^4$ | 0 | $1\times10^3$ | $6.5\times10^{-5}$ | $1.6\times10^4$ | $1.0\times10^{-3}$ | $1.0\times10^{-3}$ | $1.6\times10^2$ |
| Quadrupoles | 0.286 | $1\times10^{-3}$ | $1.06\times10^3$ | 0 | $1\times10^3$ | $9.4\times10^{-4}$ | $2.9\times10^2$ | $2.7\times10^{-4}$ | $5.6\times10^{-4}$ | $5.9\times10^2$ |

The simple first-order model shown in Fig. 1 does not describe any stray capacitances, which can be particularly significant for large superconducting magnets. An example is illustrated in Fig. 4, which shows the measured frequency response (green line) of the LHC ATLAS experiment's toroidal magnet circuit. The circuit does not have a parallel resistor ($R_p = \infty$) but the measured response only follows the model (red line) given by Eq. (4) up to about 2 mHz. It then diverges from the first-order attenuation of 20 dB/decade and follows a slower attenuation of around 6 dB/decade. This is because of unmodelled stray capacitance between the huge magnet coils. The blue line shows the theoretical response if the circuit had a 2.5 Ω parallel resistor. This matches the measured response quite well up to about 30 mHz. By defining a fictitious 2.5 Ω parallel resistor in the circuit model, the RST regulator improves the bandwidth for the rejection of perturbations by an order of magnitude.

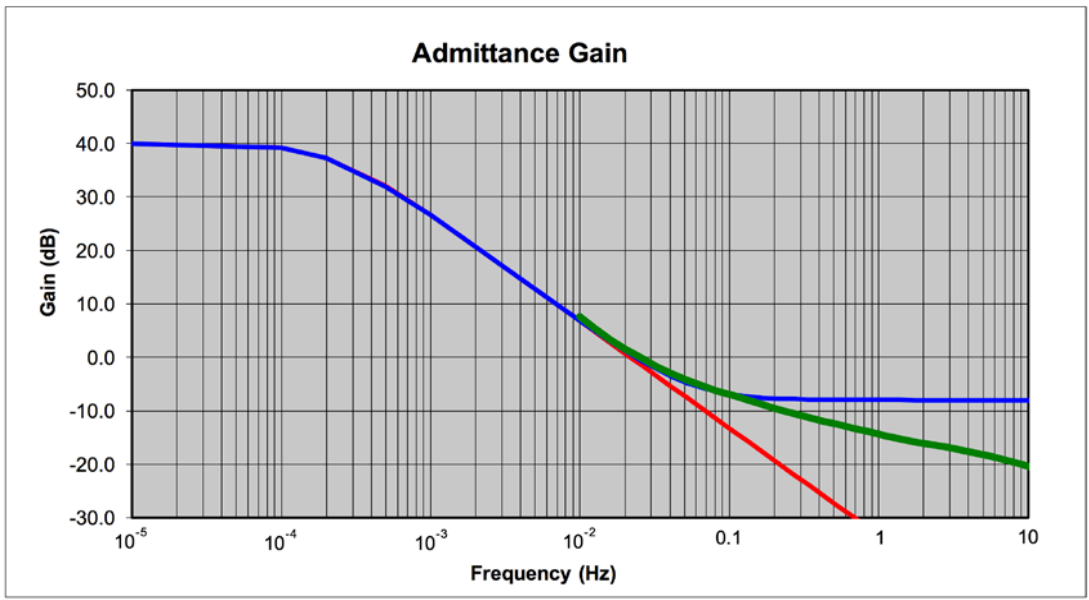

(a)

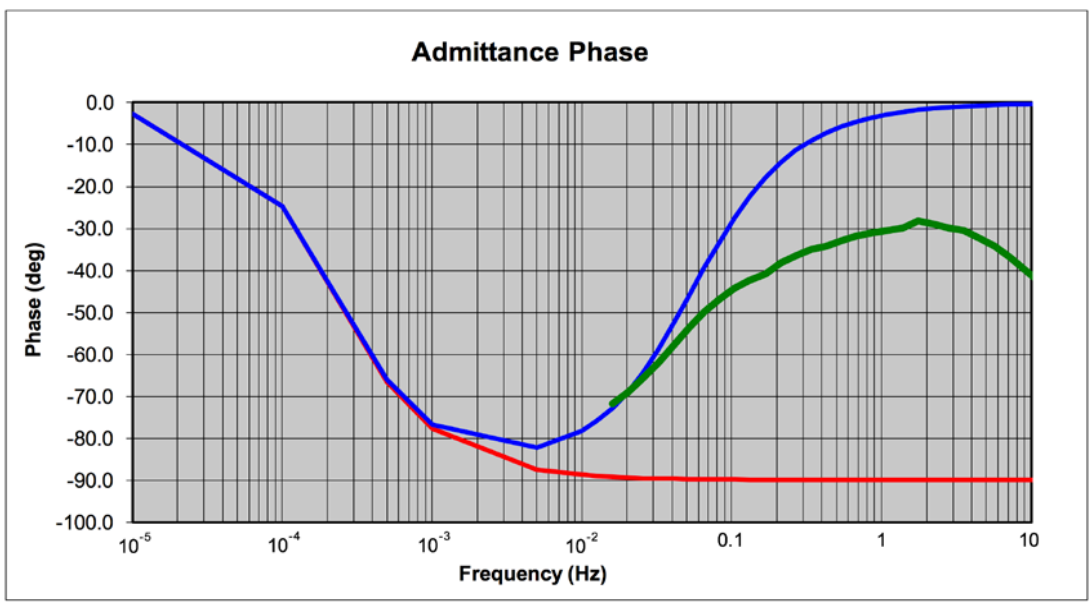

(b)

**Fig. 4:** The ATLAS experiment at the CERN LHC has a large toroidal magnet assembly comprising ten coils. Figure (a) shows the Admittance Gain (dB) and figure (b) shows the Admittance Phase (degrees) against frequency. The green lines shows the measurement. The red line shows the theoretical response of the first-order model, while the blue line shows the theoretical response of the model if the circuit had a 2.5 Ω parallel resistance ($R_p$).

### 4.3.2    Unmodelled effects

As well as stray capacitance (as mentioned above), two other unmodelled effects may be significant in some magnet circuits.

– Eddy currents. These can be particularly important for magnets that have a solid iron yoke, where the time constant of the decay of the eddy currents can be in the order of seconds. Solid yoke magnets are typically cheaper to build, so for DC applications solid magnets with slow eddy current decay time constants may be chosen. The effect of these eddy currents can be modelled as coupled inductor circuits, which can then be used to extend the load model to higher orders. However, this is not usually done and adequate performance is generally possible without it, provided that the magnet is not ramped too rapidly. For fast-cycling magnets, it is more or less essential for the magnet designers to use a laminated iron yoke that will have smaller eddy currents and a much shorter eddy current decay time constant.

– Saturation of the iron. If the magnet uses an iron yoke and if the magnetic field exceeds about 1 T, then the iron will start to saturate. When this happens, the differential inductance falls and this reduces the time constant of the circuit. This may need to be compensated, especially if the circuit is ramped rapidly. One approach to this compensation is discussed below. For DC circuits, if the saturation is not too extreme (<~20%), then it may be possible to tune the circuit for the middle value of the inductance and accept that the performance will not be optimal at the extremes.

### 4.4 Magnet saturation

Figure 5 shows a measurement of the differential inductance and stored energy for a real magnet circuit, in this case the 101 main magnets of the CERN PS accelerator. This shows the dramatic reduction in the differential inductance as the current rises due to the saturation of the iron in the yokes. When regulating the current, the time constant of the circuit will drop by more than 50%, which can lead to instability if it is not compensated. Figure 6 shows a photograph of the first PS magnet, which was produced in 1956. Figure 7 shows the measurement of the field and current for a cycle of the PS lasting 1.8 s. The field was being regulated up to 1.25 T (12 500 G), and the influence of the saturation of the iron yoke is clearly visible in the shape of the current signal as it approaches the maximum value of 5390 A. The current accelerates even as the field is decelerating.

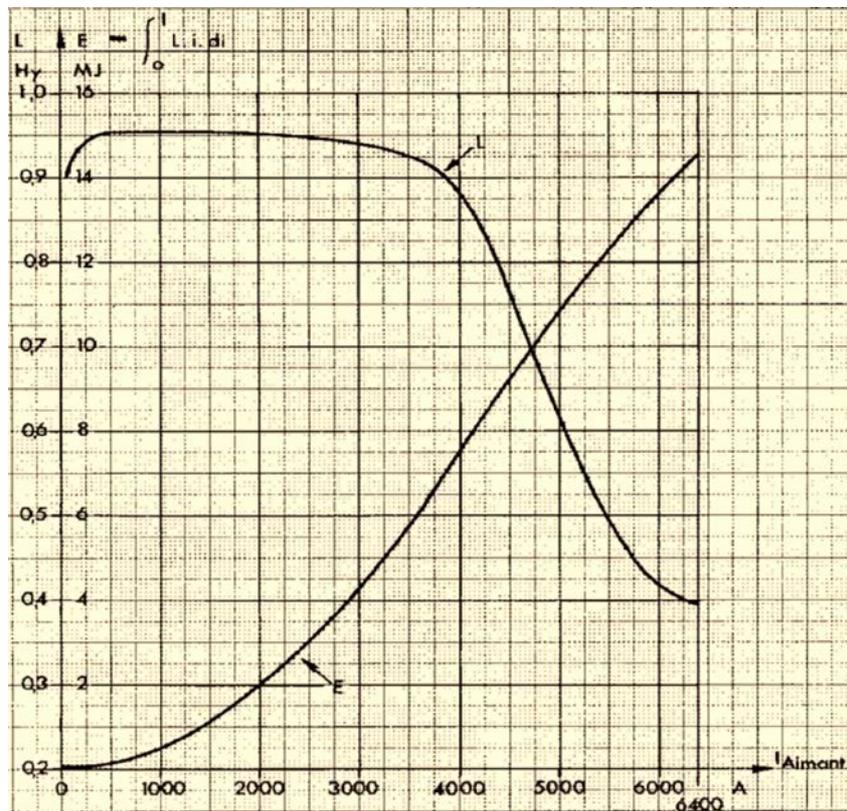

**Fig. 5:** Measurement of the differential inductance ($L$) and stored energy ($E$) of the 101 main magnets in the CERN PS accelerator, as a function of current. This illustrates the 55% reduction in the differential inductance, which the regulator must accommodate.

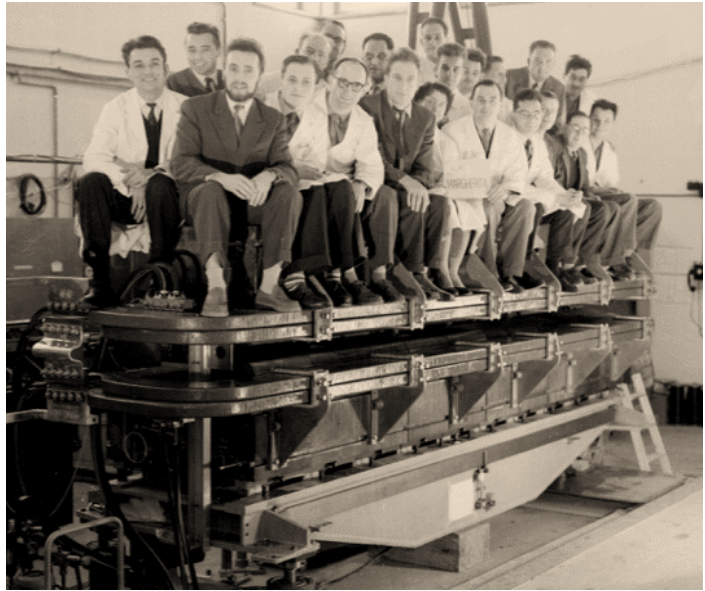

**Fig. 6:** Photograph from 1956 of the first PS magnet and members of the group who designed it. The CERN PS uses 100 of these magnets in the accelerator and one more on the surface for magnetic field measurements.

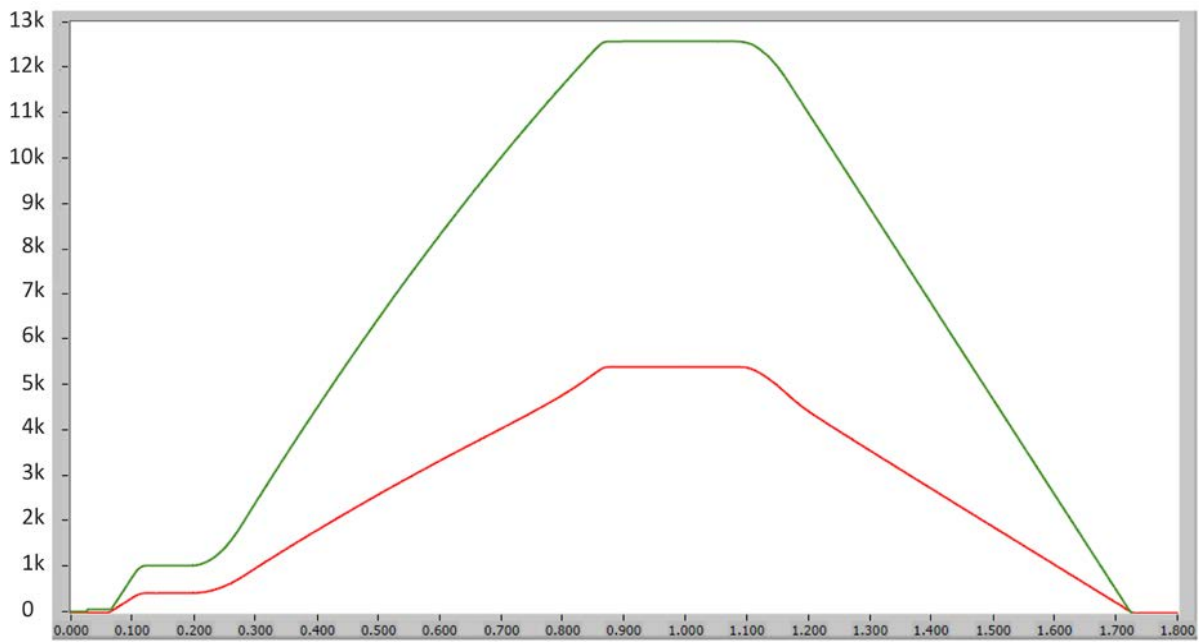

**Fig. 7:** Measurement of the magnetic field (green line in gauss) and circuit current (red line in amps) in the PS main magnets as a function of time in seconds. The magnetic field measurement is being regulated. The effect of the saturation of the iron yokes is visible in the form of the current, which accelerates at high field, even as the field is decelerating.

When regulating the magnetic field, the influence of the saturation is only a second-order effect, which can be neglected. This is the normal operating mode for the PS main magnet circuit; however, the regulator can switch between current and field regulation from cycle to cycle, and when current regulation is active, it must accommodate the change in the inductance.

Therefore, for current regulation, various strategies are possible.

– Ignore the effect of the saturation simply by reducing the bandwidth of the regulator to maintain stability even with the worst-case mismatch between the time constant of the circuit and the time constant expected by the regulator.

– Adjust the regulator parameters as a function of current. This is feasible for a classical PID regulator, but impractical for the RST algorithm, because it can be very time-consuming to calculate the RST coefficients and may even require offline computation using MATLAB (or equivalent).

– Hide the change in the inductance from the regulator by adjusting the voltage reference.

The third option is operating successfully at CERN using a surprisingly simple linear representation of the magnet inductance $L_m(I)$, as shown by the green line in Fig. 8. In this model, four parameters are used to characterize the inductance as a function of current: $L$, $L_{sat}$, $I_{sat\_start}$ and $I_{sat\_end}$. Coincidentally, the form of this linear model (in green) is the same as the gain response on the Bode diagram shown in Fig. 2. Obviously, they are completely unrelated but it can be a source of confusion.

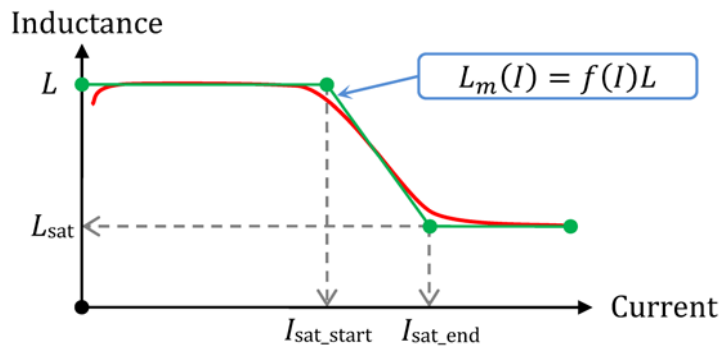

**Fig. 8:** Linear model for the reduction in the magnet inductance

The model is used to transform $V_{ref}$ from the regulation algorithm into $V_{ref\_sat}$ that is sent to the voltage source (after limitation), in order to hide the saturation effect from the regulator. This is illustrated in Fig 9.

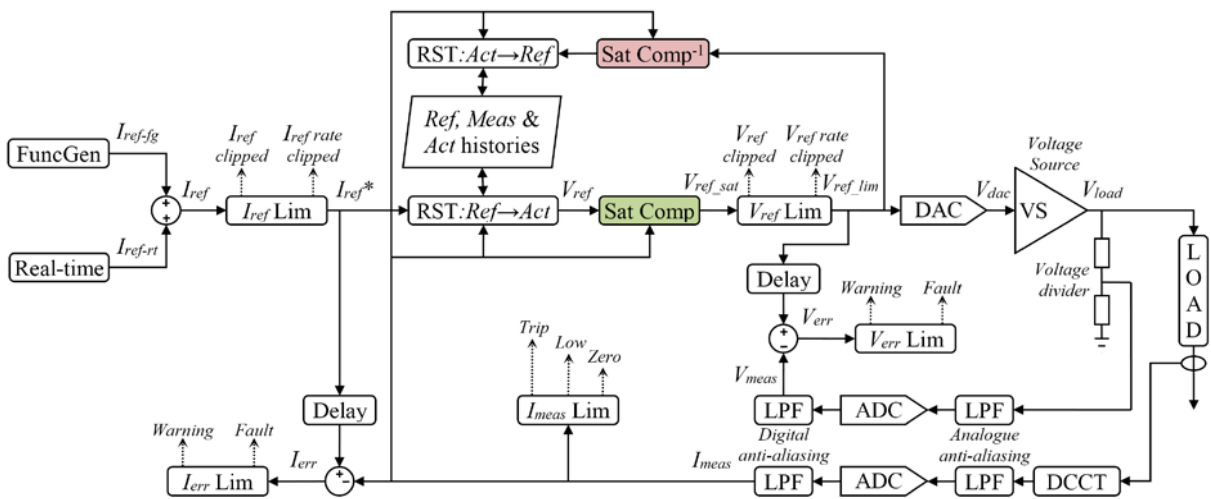

**Fig. 9:** Architecture of RST-based current regulation including saturation compensation (SatComp)

It is simple to show that $V_{\text{ref\_sat}}$ can be calculated from $V_{\text{ref}}$, $I$, and $f(I)$,

$$V_{\text{ref\_sat}} = \{1 - f(I)\}IR + f(I)V_{\text{ref}} \, , \qquad (9)$$

where $f(I) = L_m(I)/L$ and $R$ is the load resistance associated with DC operation,

$$R = R_s + \frac{R_p R_m}{R_p + R_m} \, . \qquad (10)$$

Note that as the magnet saturates, $f(I)$ reduces from 1 and Eq. (9) mixes progressively less of $V_{\text{ref}}$ and more of $IR$ into $V_{\text{ref\_sat}}$. This reduces the influence of the feedback regulator and increases the feedforward contribution based on Ohms law. If $f(I)$ becomes too small, the performance of the regulator will be compromised. Experience has shown that regulation is still effective with $f(I)$ as low as 0.5.

Figure 9 shows how $V_{\text{ref\_sat}}$ passes through a limitation block that applies minimum, maximum and rate of change limits on the reference. The output, $V_{\text{ref\_lim}}$, is transmitted to the voltage source. In this case, it is assumed to have an analogue interface, so a digital–analogue converter (DAC) is used, but it could also be via a digital serial link if the voltage source requires it. As mentioned in Eq. (3), if the voltage reference is limited, the current reference stored in the RST history must be back-calculated from the voltage reference that is actually used. This means that the saturation compensation must be reversed, which is easy,

$$V_{\text{ref}} = \frac{V_{\text{ref\_sat}} - \{1 - f(I)\}IR}{f(I)} \, . \qquad (11)$$

## 5    Controller integration

The quality of the integration of a converter controller into the wider accelerator control system can have a big impact on the reliability of the global system and the MTTR. In particular, effective analysis of powering failures depends upon having dependable logging and good tools for reviewing the logs.

### 5.1    Post-mortem logging

It is obviously essential to have access to important controller measurements in order to commission and optimize the regulation of a circuit. Furthermore, when a power converter trips, it can accelerate the analysis of the cause of the trip if the controller provides a log of the signals and events that occurred just before and after the time of the trip. This 'post-mortem' log can have two types of data:

- time series logs of important signals used in the regulation;
- time-stamped event data.

Both can be based on circular buffers. The duration of the time series logs will depend on the length of the log and the period of the sampling, while the duration of the event data will depend on how many events occurred during the period leading up to the trip. Figure 10 shows an example of an event data log from a controller in the LHC. It includes changes in state variables, diagnostics from the power converter, and commands from the upper levels of the control system.

**Fig. 10:** Example of a post-mortem event log from the CERN LHC showing the trip of a main dipole circuit. The background grey bands indicate events that occurred simultaneously.

## 5.2 Simulation of the converter and circuit

No matter which approach is taken to regulation, it is a big advantage if the controller's software includes a simulation mode in which the power converter and load are simulated in real time. This allows upper levels of the control system to be developed and tested without needing to power the circuits. It also allows other parts of the controller software to be developed and tested without a power converter.

The simulation must be simple enough to execute in real time at the iteration rate of the controller's processor. If the circuit being modelled is basically resistive, then this iteration rate might be too slow to capture the dynamics of the load. The system is effectively under-sampled. When simulating offline, for example with MATLAB Simulink, the software can simply reduce the sampling period, but this is not possible for real-time software and the program must gracefully switch from a mode in which the dynamics of the load are modelled to a mode in which they are not. The same applies to the model of the voltage source, which may have a bandwidth that is too high to be modelled by the simulation.

The challenge of including a simulation mode is not completely trivial but it is absolutely worth the effort. Even a rudimentary simulation will be helpful.

## 6 Converter control libraries

Many of the issues related to RST regulation are treated in the CERN Converter Controls Libraries, which are open source and can be used freely. The website for the libraries is https://cern.ch/cclibs.

## Acknowledgements

The author would like to thank S. Page (CERN BE-CO) and all the members of the CERN Electrical Power Converters group for contributing to the work described in this paper.

## References

[1] M. Cerqueira Bastos, *High precision current measurement for power converters*, these proceedings.

[2] B. Todd, *Radiation Risks & Mitigation in Electronic Systems*, these proceedings.

[3] F. Jenni, L. Tanner and M. Horvat, *A novel control concept for highest precision accelerator power supplies*. Proc. 10th International Power Electronics and Motion Control Conference, Cavtat and Dubrovnik, Croatia, 2002, vol. 237.

[4] I.D. Landau, *System Identification and Control Design*, (Prentice-Hall International, Englewood Cliffs, 1990).

[5] F. Bordry and H. Thiesen, *RST digital algorithm for controlling the LHC magnet current*, CERN LHC-Project-Report-258 (1999).

## Bibliography

S. McConnell and J. Detlef, *Code Complete* (Microsoft Press, Redmond, 2004), Vol. 2.

R.L. Glass, *Facts and Fallacies of Software Engineering* (Addison-Wesley Professional, Boston, 2002).

P. Koopman, *Better Embedded System Software* (Drumnadrochit Education, 2010).

E. White, *Making Embedded Systems* (O'Reilly Media, Sebastopol, 2011).

M. Barr, *Programming Embedded Systems in C and C++* (O'Reilly Media, Sebastopol, 1999).

D.E. Simon, *An Embedded Software Primer* (Addison-Wesley Professional, Boston, 1999).

J.J. Labrosse, *μC/OS-II: The Real-Time Kernel* (CRC Press, London, 2002).

P. van der Linden, *Expert C Programming: Deep C Secrets* (Prentice Hall, New Jersey, 1994).

P. O'Connor, *Practical Reliability Engineering* (Wiley-Blackwell, Hoboken, 2012).